\documentclass[manuscript]{aastex}

\usepackage{graphicx}
\usepackage{subfigure}
\usepackage{xcolor}
\usepackage{hyperref}
\usepackage{rotating}
\usepackage{tikz}
\usepackage{adjustbox}
\usepackage{blindtext}
\usepackage{multirow}

\slugcomment{}

\shorttitle{Highest frequency detection of FRB\,121102 at 
4--8 GHz using the Breakthrough Listen}
\shortauthors{Gajjar et al.}

\begin{document}


\title{Highest-frequency detection of FRB\,121102 at 4$-$8 GHz using the 
Breakthrough Listen Digital Backend at the Green Bank Telescope}


\author{V. Gajjar\altaffilmark{1},
A.~P.~V.~Siemion\altaffilmark{2,3,4,5},
D.~C.~Price\altaffilmark{2,6},
C.~J.~Law\altaffilmark{2,7},
D.~Michilli\altaffilmark{8,9},
J.~W.~T.~Hessels\altaffilmark{8,9},
S.~Chatterjee\altaffilmark{10},
A.~M.~Archibald\altaffilmark{9,8},
G.~C.~Bower\altaffilmark{11},
C.~Brinkman\altaffilmark{12},
S.~Burke-Spolaor\altaffilmark{13,14},
J.~M.~Cordes\altaffilmark{10},
S.~Croft\altaffilmark{2},
J.~Emilio Enriquez\altaffilmark{2,3},
G.~Foster\altaffilmark{2,15},
N.~Gizani\altaffilmark{2,16},
G.~Hellbourg\altaffilmark{2},
H.~Isaacson\altaffilmark{2},
V.~M.~Kaspi\altaffilmark{17},
T.~J.~W.~Lazio\altaffilmark{18},
M.~Lebofsky\altaffilmark{2},
R.~S.~Lynch\altaffilmark{19,14},
D.~MacMahon\altaffilmark{2},
M.~A.~McLaughlin\altaffilmark{13,14},
S.~M.~Ransom\altaffilmark{20},
P.~Scholz\altaffilmark{21},
A.~Seymour\altaffilmark{22,19},
L.~G.~Spitler\altaffilmark{23},
S.~P.~Tendulkar\altaffilmark{17},
D.~Werthimer\altaffilmark{1,2},
Y.~G.~Zhang\altaffilmark{2}
}
\small
\altaffiltext{1}{Space Sciences Laboratory, 7 Gauss way, University of California, Berkeley, CA, 94720, USA}
\altaffiltext{2}{Department of Astronomy, University of California, Berkeley, 501 Campbell Hall \#3411, Berkeley, CA, 94720, USA}
\altaffiltext{3}{Radboud University, Nijmegen, Comeniuslaan 4, 6525 HP Nijmegen, The Netherlands}
\altaffiltext{4}{SETI Institute, Mountain View, California}
\altaffiltext{5}{Institute of Space Sciences and Astronomy, University of Malta}
\altaffiltext{6}{Centre for Astrophysics \& Supercomputing, Swinburne University of Technology, Hawthorn, VIC 3122, Australia}
\altaffiltext{7}{Radio Astronomy Lab, University of California, Berkeley, CA, 94720, USA}
\altaffiltext{8}{ASTRON, Netherlands Institute for Radio Astronomy, Postbus 2, 7990 AA, Dwingeloo, The Netherlands}
\altaffiltext{9}{Anton Pannekoek Institute for Astronomy, University of
Amsterdam, Science Park 904, 1098 XH Amsterdam, The Netherlands}
\altaffiltext{10}{Cornell Center for Astrophysics and Planetary Science and Department of Astronomy, Cornell University, Ithaca, NY 14853, USA}   \altaffiltext{11}{Academia Sinica Institute of Astronomy and Astrophysics, 645 N. A'ohoku Place, Hilo, HI 96720, USA}
\altaffiltext{12}{Physics Department, University of Vermont, Burlington, VT 05401, USA}
\altaffiltext{13}{Department of Physics and Astronomy, West Virginia University, Morgantown, WV 26506, USA}
\altaffiltext{14}{Center for Gravitational Waves and Cosmology, Chestnut Ridge Research Building, Morgantown, WV 26505, USA}
\altaffiltext{15}{University of Oxford, Sub-Department of Astrophysics, Denys Wilkinson Building, Keble Road, Oxford, OX1 3RH, United Kingdom}
\altaffiltext{16}{Hellenic Open University, School of Science and Technology, Parodos Aristotelous 18, 26 335 Patra, Greece}
\altaffiltext{17}{Department of Physics and McGill Space Institute, McGill University, 3600 University, Montr\'{e}al, QC H3A 2T8, Canada}
\altaffiltext{18}{Jet Propulsion Laboratory, California Institute of Technology, Pasadena, CA 91109, USA}
\altaffiltext{19}{Green Bank Observatory, PO Box 2, Green Bank, WV 24944, USA}
\altaffiltext{20}{National Radio Astronomy Observatory, Charlottesville, VA 22903, USA}
\altaffiltext{21}{National Research Council of Canada, Herzberg Astronomy and Astrophysics, Dominion Radio Astrophysical Observatory, P.O. Box 248, Penticton, BC V2A 6J9, Canada}
\altaffiltext{22}{National Astronomy and Ionosphere Center, Arecibo Observatory, PR 00612, USA}
\altaffiltext{23}{Max-Planck-Institut f\"{u}r Radioastronomie, Auf dem H\"{u}gel 69, D-53121 Bonn, Germany}

\normalsize


\begin{abstract}
We report the first detections of the repeating fast radio burst source FRB\,121102
above 5.2~GHz. Observations were performed using the 4$-$8~GHz receiver of the
Robert C. Byrd Green Bank Telescope with the Breakthrough Listen digital
backend. We present the spectral, temporal and polarization properties of 21 bursts
detected within the first 60 minutes of a total 6-hour observations. 
These observations comprise the highest burst density yet reported in the literature, 
with 18 bursts being detected in the first 30 minutes.  A few bursts clearly show temporal 
sub-structures with distinct spectral properties. These sub-structures superimpose to 
provide enhanced peak signal-to-noise ratio at higher trial dispersion measures. 
Broad features occur in $\sim 1$~GHz wide subbands that typically differ in peak frequency 
between bursts within the band. Finer-scale structures ($\sim 10-50$~MHz) within these 
bursts are consistent with that expected from Galactic diffractive 
interstellar scintillation. The bursts exhibit nearly 100\% linear 
polarization, and a large average rotation measure of 
9.359$\pm$0.012~$\times$~10$^{\rm 4}$~rad~m$^{\rm -2}$ 
(in the observer's frame). No circular polarization was found for any burst. 
We measure an approximately constant polarization position angle in the
13 brightest bursts. The peak flux densities of 
the reported bursts have average values (0.2$\pm$0.1~Jy), 
similar to those seen at lower frequencies ($<3$~GHz), while the average burst widths 
(0.64$\pm$0.46 ms) are relatively narrower. 
\end{abstract}





\section{Introduction} 

Fast Radio Bursts (FRBs) are a class of radio transients with inferred
extragalactic origin due to their anomalously high dispersion measures (DMs)
relative to the contribution expected from the Galactic electron distribution \citep{Lor07}. 
This inference was proven for one source, FRB\,121102, when repeated bursts at the DM of 
557 pc~cm$^{\rm -3}$ were localized by interferometry to be unambiguously associated 
with a dwarf galaxy at redshift $z~ = 0.193$ \citep{Cha17,Mar17,Ten17}. Bursts from FRB\,121102 
have isotropic apparent radio energies of $10^{40}$~erg, several orders of magnitude 
higher than for any other radio transient on millisecond timescales \citep{Law17}.

FRB\,121102 is the only FRB known to repeat and for which a position is known to sub-arcsecond precision. This has facilitated extensive follow-up observational campaigns. While the  burst cadence is irregular, there appear to be epochs in which the source is more active and multiple bursts are detected. 
For example, \cite{Sch16} reported 6 bursts within a 10-minute interval while several other long observing sessions resulted in non-detections (e.g. \citealt{pgr18}). These bursts have so far only been detected between $1 - 5.2$\,GHz (\citealt{Spi16,Sch16,Law17,Michilli:2018}; Spitler et al. 2018 in prep). 
Additionally, \cite{Law17} reported the non-detection of bursts at 70\,MHz, 4.5\,GHz, and 15\,GHz, 
during epochs in which bursts were detected at 1.4 and 3\,GHz, 
indicating that some bursts are not likely to be broadband. 

Propagation effects, such as scintillation and plasma lensing, can significantly alter the 
observed radio emission from impulsive radio sources \citep{2015MNRAS.451.3278M,Cor17}.  
Estimates of the FRB burst rate have taken into account the observed bursts 
(e.g. \citealt{2017AJ....154..117L}); however the observed radio 
emission could be affected by these propagation effects. 
Disentangling the intrinsic emission from propagation effects is thus an important goal in FRB science.
Detection of bursts at different frequencies and over wider bandwidths is helpful for studying the frequency dependence of potential propagation effects. Measurement of burst polarization can also shed light on emission physics and the source's local environment.  Recently, \cite{Michilli:2018} reported a very high and variable Faraday rotation measure of ${\sim}10^{\rm 5}$ rad m$^{\rm -2}$ for FRB\,121102, suggesting that this source is embedded in an extreme and dynamic magneto-ionic environment. 

Here we report the detection of 21 bursts from FRB\,121102---all of which
occurred within an hour---using the $4 - 8$\,GHz receiver on the Robert C. Byrd Green 
Bank Telescope (GBT) and the Breakthrough Listen backend. These are the 
highest-frequency detections of bursts from any FRB to date\footnote{Fifteen of these detections were announced briefly in \cite{gsm+17}. Here we are reporting more detailed analysis.}. 

The remainder of this paper is structured as follows.  In Section \ref{sect:observations}, we describe our observations. In Section \ref{sect:dm_optimize} we highlight issues with finding a true DM for the bursts, followed by a discussion on the spectro-temporal structures in a subset of bursts in Section \ref{sec:spectro-temp}. We confirm and extend the results of \cite{Michilli:2018} by performing polarimetry on baseband voltage data in Section \ref{sect:flux_and_pol_analysis}. Our wide instantaneous bandwidth reveals these bursts to be band-limited, with dynamic structure varying on short timescales, which are further discussed in Section \ref{sect:b2b_analysis}. A brief discussion of our findings is in Section \ref{sect:discussion} along with a summary in Section \ref{sect:conclusion}. 
 
\section{Observations} 
\label{sect:observations}
Observations of FRB\,121102 were conducted with the GBT as part of the
Breakthrough Listen (BL) project \citep{wds+17}. BL is a comprehensive search
for extraterrestrial intelligence (SETI) employing both optical and radio
telescopes; the BL target list includes nearby stars and nearby galaxies, as well
as other anomalous astronomical sources broadly classified as ``exotica'' \citep{ism+17}. 
As a component of the latter category, BL is conducting a targeted SETI 
search towards known FRB positions to investigate any associated 
artificial signals and/or underlying modulation pattern, hypothesizing that 
one or more FRBs may be deliberate artificial beacons or other manifestations of technology (e.g., \citealt{lingam2017})\footnote{However, we emphasize here that it is unlikely that the bursts we detected were transmitted from an intelligent civilization.}. 

Observations using the $4 -8$\,GHz (C-band) receiver on the GBT were conducted on 
2017 August 26 during a 6-hour BL observing block. The initial hour (with scans 
numbered from 0 to 10) was used for configuration of various telescope settings 
and calibration procedures.  The calibrator 3C\,161 and an off-source position 
were observed for one minute each, along with a calibration noise diode, and 
used for flux and polarization calibration using the  \texttt{pac} tool in 
\texttt{PSRCHIVE}. A 5-minute observation of the bright pulsar PSR\,B0329+54 
was also performed as a diagnostic to verify polarization and flux density 
measurements.  The remaining five hours of the session were divided into 
ten 30-minute scans of FRB\,121102, identified with scan numbers ranging from 11 to 20.  

Observations were conducted with the BL digital backend \citep{mpl+17}, which
recorded 8-bit baseband voltage data across the entire 4\,GHz receiver bandwidth.
The GBT analog downconversion system provided four 1500\,MHz wide tunable passbands 
to cover the $4 - 8$\,GHz band, configured with central on-sky
frequencies of 4625.0, 5937.5, 7250.0, and 8562.5\,MHz. These frequencies were
chosen such that the signal at the edges of each passband's intermediate frequency
(IF) filter overlapped the adjacent passband by 187.5\,MHz. 

Each dual-polarization passband was Nyquist-sampled using 8-bit digitizers, polyphase channelized to 512 `coarse' frequency channels, requantized to 8 bits, and then
distributed to a cluster of compute nodes that recorded these data to disk.
Directly after observations, the coarsely-channelized voltage data were further
channelized (to 366\,kHz resolution), integrated
(with a sampling time of 350\,$\mu$sec) using a custom GPU-accelerated spectroscopy code, 
and finally written to\texttt{SIGPROC}\footnote{http://sigproc.sourceforge.net} filterbank files as
total intensity (Stokes I) dynamic spectra with 4096 spectral channels. 

These dynamic spectra were searched using the \texttt{Heimdall} package
\citep{bbb+12} for dispersed pulses within the DM range of 500 to
700~pc~cm$^{\rm -3}$, using a DM interval of 0.1~pc~cm$^{\rm -3}$. 
We detected 21 bursts above a threshold signal-to-noise ratio (S/N) of six. 
A section of raw voltages (of total 1.5~seconds) around each detected burst 
was extracted for further processing and coherently dedispersed to a DM of 
565.0~pc~cm$^{\rm -3}$  (see Section \ref{sect:dm_optimize}) using the 
\texttt{DSPSR} package \citep{vb11}.  
The coherently dedispersed \texttt{PSRFITS} data products have temporal and spectral 
resolutions of 10.1\,$\mu$s and 183\,kHz, 
respectively\footnote{PSRFITS data for all 
these bursts are publicly available to download 
at \url{seti.berkeley.edu/frb121102/}}. We have 
discarded lower part of the frequency band (4 to 4.5 GHz) 
in further analysis due to spurious radio frequency interferences. 

\begin{table*}[h]
\tiny
\begin{center}
\begin{tabular}{l c c c c c c c c c}
\hline
Pulse & MJD (57991+) & S/N  & DM$_{\rm S/N}$ (pc~cm$^{\rm -3}$) & S$_{\rm peak}$ (mJy) & Width (ms) & F(Jy~ms) & RM$_{\rm obs}$ (rad~m$^{\rm -2}$) & PA$_{\rm \infty}^{\rm mean}$ (deg) & $\Delta$PA$_{\rm \infty}$ (deg) \\
\hline
\hline 
11A &  0.409904044  & 55.7  & 601$\pm$18 & 380.5 & 1.74$\pm$0.01 & 0.62 & 93559$\pm$18 &  63.9$\pm$0.4 & 6	\\ 
11B  & 0.412764720 & 7.1   & 587$\pm$4  & 51.9 & 0.39$\pm$0.16  & 0.02  & $-$          &  $-$ 		  & $-$ \\ 
11C  & 0.413019871 & 9.9   & 574$\pm$15 & 85.2 & 0.42$\pm$0.08  & 0.03  & $-$          &  $-$ 		  & $-$ \\ 
11D  & 0.413458764& 37.0  & 583$\pm$33 & 314.8 & 0.68$\pm$0.08 & 0.23 & 93503$\pm$53 &  79.2$\pm$0.3 & 4 \\
11E  & 0.413706653 & 28.2  & 591$\pm$18 & 126.8 & 1.15$\pm$0.08 & 0.14 & 93577$\pm$43 &  80.9$\pm$0.8 & 6 \\
11F  & 0.413837058 & 29.2  & 588$\pm$49 & 157.5 & 0.92$\pm$0.08 & 0.13 & 93836$\pm$160&  86.5$\pm$0.7 & 5 \\
11G  & 0.416436793 & 7.6   & 574$\pm$10 & 52.9 & 0.31$\pm$0.08 	& 0.02 & $-$ 		  &  $-$		  & $-$ \\
11H  & 0.416633362 & 58.0  & 575$\pm$33 & 699.9 & 0.27$\pm$0.01	& 0.18 & 93467$\pm$35 &	 76.4$\pm$0.3 & 8 \\
11I  & 0.417714722 & 13.4  & 569$\pm$10 & 125.5 & 0.34$\pm$0.08	& 0.04 & 93648$\pm$59 &  77.2$\pm$1.2 & 4 \\
11J  & 0.417865553 & 9.9   & 573$\pm$19 & 118.6 & 0.39$\pm$0.08	& 0.04 & $-$ 		  &  $-$		  & $-$ \\
11K  & 0.418627200 & 13.9  & 591$\pm$32 & 105.2& 0.81$\pm$0.08 	& 0.06 & 93567$\pm$248 &  77.9$\pm$1.3 & 5 \\
11L  & 0.419449885 & 8.4   & 590$\pm$5  & $-$  & $<$0.70 		& $-$ &  $-$ 		  & $-$ 		  & $-$ \\
11M  & 0.421212904 & 11.1  & 560$\pm$6  & 94.7 & 0.41$\pm$0.08 	& 0.04 & $-$ 		  & $-$ 		  & $-$ \\
11N  & 0.421712667 & 12.4 & 567$\pm$7  & 256.7 & 0.19$\pm$0.01 	& 0.04 & 93513$\pm$60  & 81.8$\pm$0.8  & 5 \\
11O  & 0.422939456 & 12.1 & 584$\pm$14 & 139.9 & 0.41$\pm$0.08 	& 0.04 & 93802$\pm$224 & 76.9$\pm$1.2  & 2 \\
11P  & 0.424270656  & 6.4  & 601$\pm$31 &$-$    & $<$1.79 	    & $-$ 		   & $-$		  &  $-$ \\
11Q  & 0.426552515 & 20.9 & 562$\pm$23 & 400.2 & 0.18$\pm$0.01 	& 0.07  & 93502$\pm$62  &79.5$\pm$0.4  &  5 \\
11R  & 0.430427904 & 7.4  & 582$\pm$8  &$-$    & $<$0.27 		& $-$ & $-$           & $-$ 		  &  $-$ \\
12A  & 0.431974007  & 13.6 & 572$\pm$56 & 132.2 & 0.72$\pm$0.08	& 0.07 & 93452$\pm$111 & 80.4$\pm$0.4 &  5 \\
12B  & 0.439360677  & 20.6 & 636$\pm$11 & 331.4 & 1.68$\pm$0.01	& 0.45 & 93643$\pm$62  & 61.8$\pm$0.6 &  7 \\
12C  & 0.448427650  & 10.7 & 580$\pm$40 & 118.8 & 0.68$\pm$0.08	& 0.07 & 93593$\pm$101 & 73.5$\pm$0.7 &  11 \\
\hline
\end{tabular}
\caption{Properties of all 21 bursts. Columns list, from left to right:
burst numbers, barycentric arrival time at infinite frequency, S/N, DM that
maximizes S/N, peak flux density, width of the burst, fluence, RM (in  the observer's frame), 
weighted average PA at infinite frequency, and standard deviation of PA across the pulse 
phase, respectively. We were not able to flux calibrate bursts 11L, 11P, and 11R due to low S/N. 
Also for these pulses, we sub-integrated adjacent time bins and hence could only obtain an upper 
limits on the widths.} 
\label{tab:detections_main}
\end{center}
\end{table*}
\normalsize

\section{Analysis}
\label{sect:analysis}

\texttt{PSRCHIVE} \citep{2012AR&T....9..237V} and custom Python scripts were
used to measure the properties of the detected bursts.  Table
\ref{tab:detections_main} lists various parameters determined for each burst.
Among the ten recorded scans, all reported detections occurred in the first
hour, i.e. from scans 11 and 12.  We assigned the bursts identifiers 11A through
11R, and 12A through 12C, according to their scan number and order of arrival.


%
%

\subsection{DM Optimization}
\label{sect:dm_optimize}
\begin{figure}
    \begin{center}
    \centering
    \includegraphics[scale=0.8]{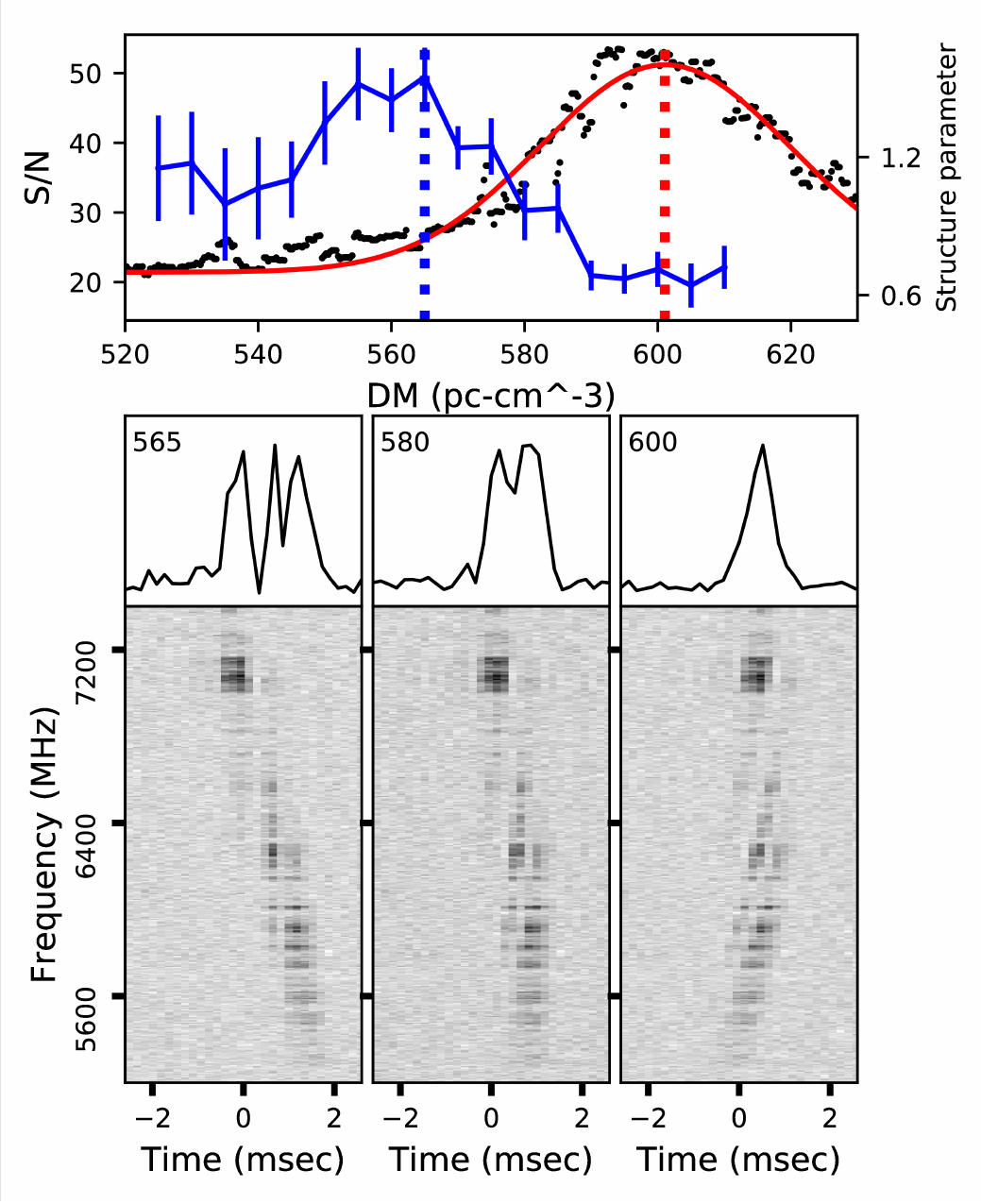}
    \caption{DM analysis for burst 11A. The top panel shows the S/N
    (black) and structure parameter (blue) as a function of DM. The black 
    points are S/N measurements and the red line is a best fit model of a Gaussian
    function.  The blue points are measured structure parameter values with
    corresponding errors. The red and blue dotted lines are for S/N and 
    structure maximizing DMs, respectively. The bottom three panels 
    show dynamic spectra along with average pulse profiles 
    at three different trial DMs, labeled in the top left 
    corners. At the highest DM trial of 600~pc~cm$^{-3}$, 
    the alignment of burst structures produces the highest S/N while at the 
    trial DM of 565~pc~cm$^{-3}$ these structures appear to be well separated.}
    \label{fig:dm_optimization}
    \end{center}
\end{figure}

The DM of FRB\,121102 has been previously reported to range between
$553 - 569$~pc~cm$^{-3}$ \citep{Spi14,Spi16, Sch16,Law17} from lower
frequency observations. For each burst detected here, we measured the S/N as a function of DM, 
and found that individual bursts show slightly different peak S/N-maximizing DMs 
(Table \ref{tab:detections_main}). To investigate further, we coherently 
dedispersed raw voltages from burst 11A across a range of DMs with a DM 
step of 5 pc~cm$^{-3}$. This investigation lead to the discovery of 
detailed spectro-temporal structures ({\itshape components}, see 
Figure \ref{fig:dm_optimization} and Section \ref{sec:spectro-temp}). 
It is likely that the differences between the DMs originate from how 
these components superimpose for different trial values of DM. 
Figure \ref{fig:dm_optimization} shows burst 11A dedispersed 
and detected at three different DMs. While it is not clear if 
these components are intrinsic to the emission mechanism or 
due to propagation effects, their peculiar alignment provides 
enhanced S/N at higher DMs compared to previously reported 
DMs \citep{Spi14,Spi16, Sch16,Law17}. 

Assuming these components to be intrinsic, we optimized the S/N of individual components by 
maximizing the average absolute rate-of-change of total flux density, which we call here a 
{\em structure parameter}, across the pulse window defined as,
\begin{equation}
{\rm Structure ~ Parameter} = \frac{1}{n}\sum_{i}^{n} \left|\frac{S_i - S_{i+1}}{\Delta{t}}\right|. 
\end{equation}
Here,  $n$ is the number of on--pulse bins while $S_i$ is the 
flux at the $i^{\rm th}$ bin and $\Delta{t}$ is the time resolution. 
This structure-parameter-maximizing DM differs significantly from the S/N-maximizing DM (see Figure \ref{fig:dm_optimization}). Similar techniques have also been explored and will be reported in detail in Hessels et al. (2018, in prep) at lower frequencies. Although an analysis of this type was not possible for most of our detected pulses due to either low S/N, fewer components, or both, we assumed a structure-maximizing DM of 565.0~pc~cm$^{-3}$ to be consistent for all our bursts and used that value in coherent dedispersion. We note that this estimate can be off by our DM resolution in Figure \ref{fig:dm_optimization} from a `true' DM (i.e. $\pm$5~pc~cm$^{-3}$). 

\subsection{Spectro-temporal components\label{sec:spectro-temp}}
\begin{figure}
\begin{adjustbox}{addcode={\begin{minipage}{\width}}{\caption{\small
    Dynamic spectra of all 21 detected bursts, coherently dedispersed at the DM of 
    565~pc~cm$^{-3}$. Waterfall data are frequency (Y-axis) vs. time (X-axis) with units 
    of MHz and ms, respectively, with intensity indicated on a linear scale. The gray 
    scales are different and scaled according to individual pulse S/N. Top 
    panels show flux density (mJy), and for twelve 
    bursts circular (blue) and linear (red) polarization and position angle (PA)
    referenced to infinite frequency. The dashed horizontal 
    blue lines are frequency boundaries as the profile for each burst was
    obtained after adding frequencies between these boundaries to enhance S/N.
    \label{fig:collage}
    }\end{minipage}},rotate=90,center}
    \includegraphics[scale=0.65]{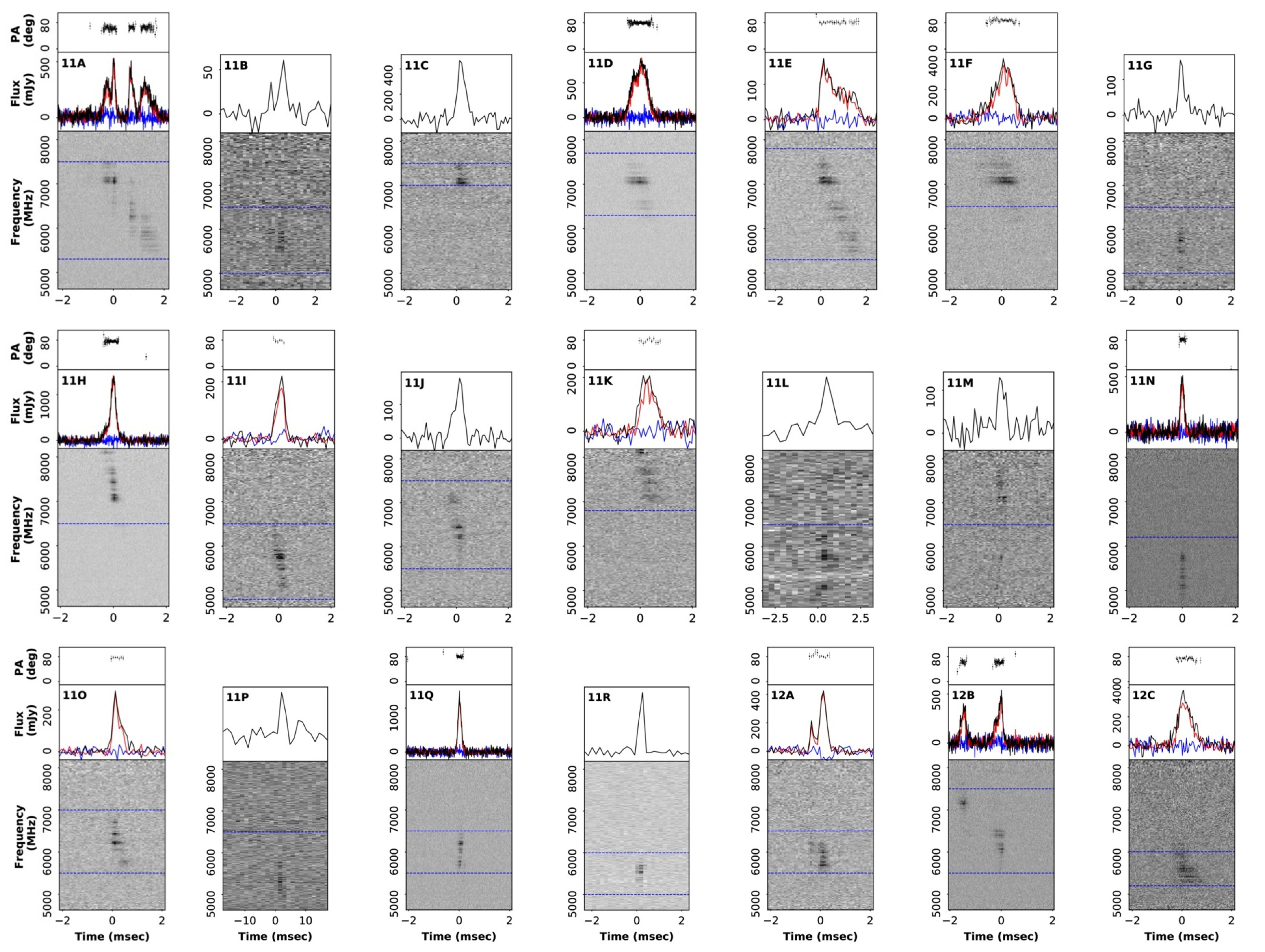}
    
\end{adjustbox}
\end{figure}

We found highly variable temporal and spectral features in many of the bursts.
Figure \ref{fig:collage} shows dynamic spectra of all bursts after coherent
dedispersion at a DM of 565~pc~cm$^{-3}$. Bursts 11A and 12B exhibit distinct
components, while bursts such as 11E, 11K, and 11O show some indication of
unresolved components. For bursts 11D and 11F, such components are not clearly
visible but likely give rise to the slanted or curved features seen in the
dynamic spectra.  Each burst was modeled as a sum of multiple Gaussian components using the
\texttt{PSRCHIVE} utility \texttt{paas}. The model was then used to derive average widths and fluences for all bursts (see Table \ref{tab:detections_main}). We note that burst widths found here are relatively narrow compared to burst
widths seen at lower frequencies (<3 GHz) - as also noted for the Arecibo sample of 4.5-GHz bursts presented by \cite{Michilli:2018}.

As 11A and 12B clearly exhibit distinct components, we measured the component
widths separately.  We identify four components in burst 11A integrated across $4.5 - 8$\,GHz, 
with individual widths of 0.18, 0.10, 0.25, and 0.32\,ms from the leading component to the trailing
component, respectively.  These components show varying spectral extent, with
the first two limited to higher frequencies and the last two at lower frequencies. 
Additionally, components three and four show a very sharp rise and gradual decay
of the burst energy.  Component four exhibits a gradual increase in burst width
towards lower frequencies. To compare with the expected scaling from a power-law 
scattering model ($\propto \nu^{-4.4}$ where $\nu$ is the frequency of observations), 
we modeled the increase in the width as a function of frequency for 
the fourth component. We extracted dynamic spectra with widths of 200\,MHz centered 
around 6200, 5980, 5780,  and 5460\,MHz. An integrated profile of the fourth component 
at these frequencies was found to have widths of 0.19, 0.49, 0.55 and 0.65\,ms, respectively. 
We did not find width variations to follow a power-law trend, which is expected from scattering. 
More detailed analyses of the progressive drift in pulse profile components, 
as seen at other radio frequency bands, are presented in Hessels et al.\ (2018, in prep).

Burst 12B also exhibits two resolved components with widths of 0.29 and 0.30\,ms,
separated by $\sim$2\,ms. As with 11A, the leading component spans
higher frequencies, while the trailing component spans lower frequencies.  In
contrast to 11A, the trailing component of 12B does not show an increase in
width at lower frequencies. While there may be a third component on the 
leading edge of the second component, the burst S/N is
not sufficient for verification.

\subsection{Faraday Rotation}
\label{sect:flux_and_pol_analysis}
We determined the Faraday rotation measure (RM) of the bursts by performing a
brute-force  search of the linear polarization fraction at different RM trials
with the \texttt{rmfit} tool from  \texttt{PSRCHIVE}. We developed a custom routine to correct 
for the initial RM values from \texttt{rmfit} and fit a quadratic function to the complex 
angles determined by the Stokes parameters. This technique refines the initial RMs 
and computes a polarization angle (PA) for each burst. A detailed description of this technique can be found in \cite{Michilli:2018}. This approach was motivated by the extremely variable spectral properties of the bursts.  From the 13 brightest bursts that could be analyzed using this technique, we find a mean RM of $93589 \pm 118$ rad m$^{-2}$. The RMs found here differ by $9169 \pm 122$ rad m$^{-2}$ from the RMs of the bursts detected from Arecibo at 4.5\,GHz by \cite{Michilli:2018}. After de-rotating the Stokes Q and U using the best-fit value of RM, all 13 bursts show $\sim 100\%$ linear polarization fraction and uniform polarization angles across the burst (Figure \ref{fig:collage}). We have only considered PAs from the phase bins with linear polarization above 3 sigma. The weighted mean PA for each burst, measured after weights calculated empirically from the variance in 
PAs across these phase bins, is listed in Table \ref{tab:detections_main}. The uncertainty 
on the weighted PA was measured as a sum of all weights for each burst. It should be noted 
that the PA is roughly uniform across the pulse phase, however, there are apparently significant 
variations in the average PA between all observed bursts at the $\sim 3 \sigma$ level. 
For example, bursts 11A and 12B show slightly lower PAs compared 
to rest of the reported bursts. These differences in PAs could possibly arise from our uncertainties 
in the RM measurements, as these two quantities are covariant. No circular polarization was found 
for any burst and any undetected circular polarization is less than a few percent. 

\subsection{Burst-to-burst spectral variation}
\label{sect:b2b_analysis}

\begin{figure*}[ht]
    \centering
    \centering
    \includegraphics[scale=0.7]{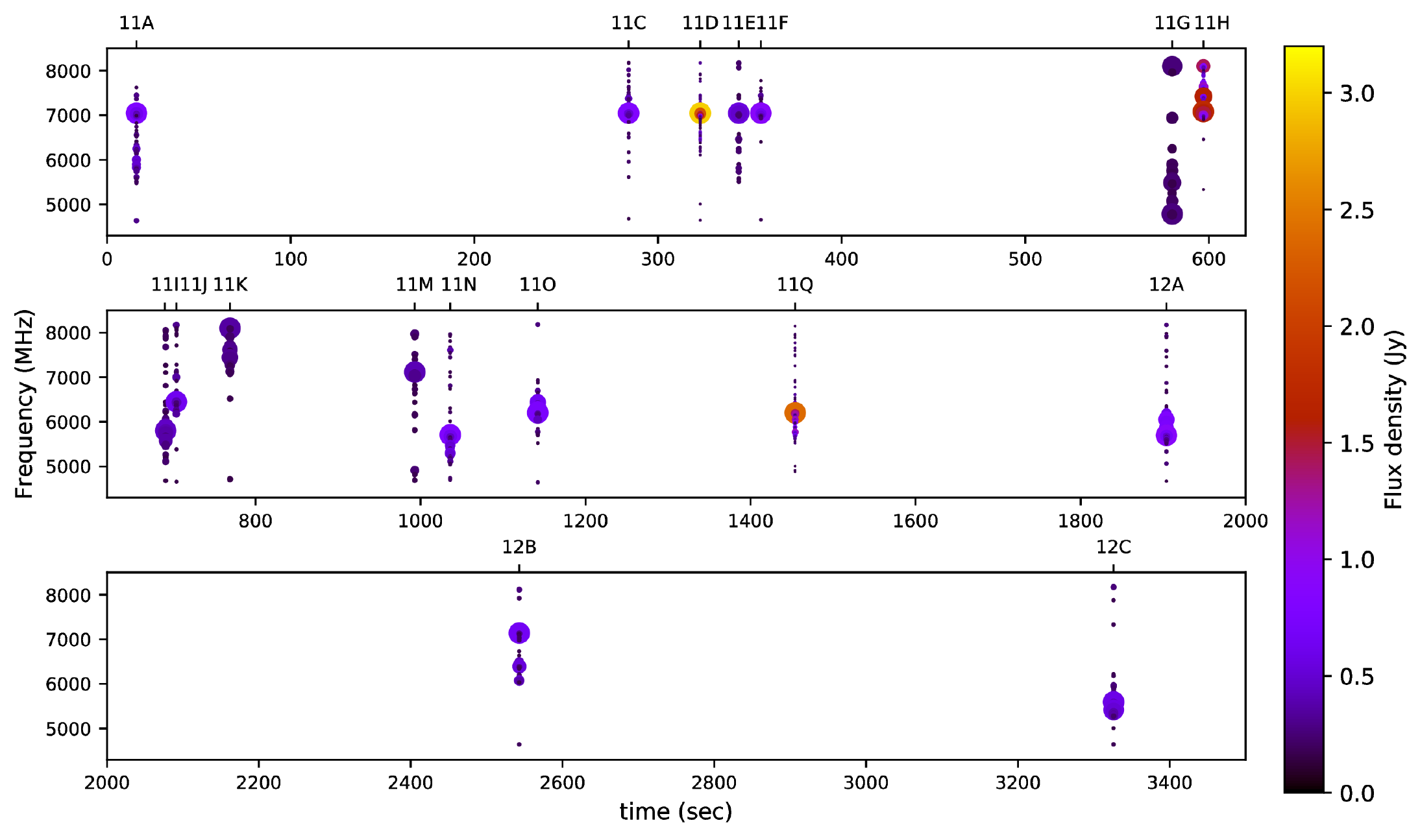}
    \caption{Burst-to-burst spectral properties as a function of time for the 
    FRB\,121102 bursts presented here. Panels (top to bottom) show spectra for each 
    burst at their corresponding arrival times for 17 bright bursts 
    detected during the first observing hour. Each spectrum is shown
    with 11\,MHz of spectral resolution to highlight large-scale frequency 
    structures and rapid variation of spectral peaks. Color represents absolute flux 
    density across all bursts while circle size represents flux density relative to peak flux density for a given
    burst on a base-2 logarithmic scale. 
    }
    \label{fig:dynamic_spectra} 
\end{figure*}

\begin{figure}[h]
    \centering
    \includegraphics[scale=0.6]{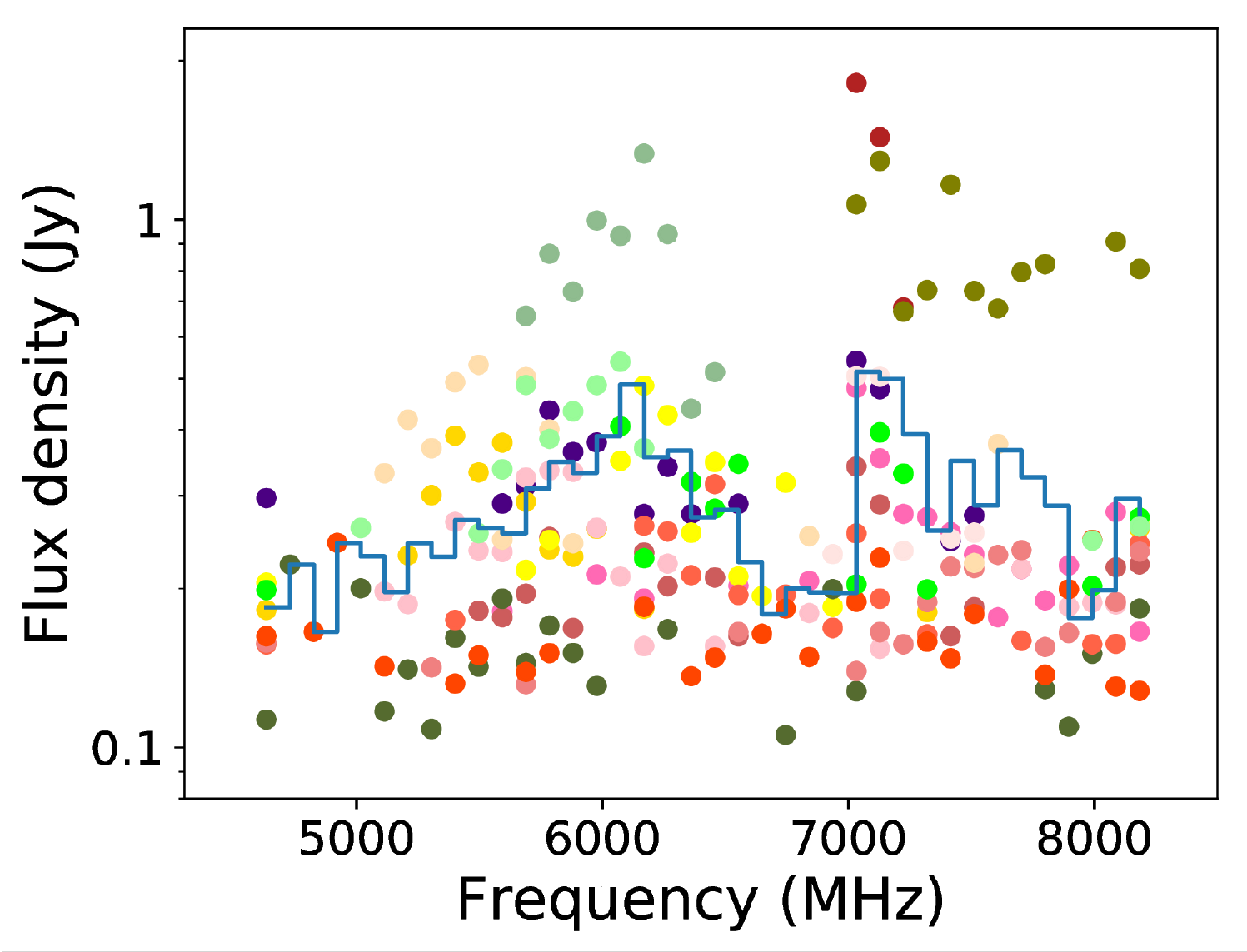}
    \caption{Average spectral energy distribution of FRB\,121102 bursts presented 
    here. Different colors represent histograms of spectral energy for different 
    bursts, binned with a resolution of 94\,MHz. The blue line represents the mean. 
    Most bursts show spectral peaks near 6\,GHz or 7.2\,GHz and an apparent trough 
    between 6.5 to 7\,GHz.}
    \label{fig:avg_spectra}
\end{figure}

Figure \ref{fig:dynamic_spectra} shows measured flux density as a function of 
observing frequency and time for the 17 brightest bursts. The peak
flux frequency is not constant within the observed frequency band; similar
spectral variation between bursts has also been reported by previous studies \citep{Spi16,Sch17,Law17,Michilli:2018}. 
These large-scale frequency features are approximately of the order of $\sim 1$\,GHz in extent. 
To estimate the average number of discrete large-scale structures, we divided the
spectrum from each burst into multiple bins, each 94\,MHz wide, and then calculated
the total amount of spectral energy in each bin (Figure \ref{fig:avg_spectra}).
The overall spectral energy distribution shows two approximately GHz-wide peaks around 6\,GHz
and 7.2\,GHz, with a possible trough in the spectral energy between 6.5 to 7\,GHz. 
It should be noted that this spectral behavior is based on our detections of only 
21 bursts. It is possible that the detection of a larger sample of bursts might fill this trough. 

Figure \ref{fig:dynamic_spectra} shows that bursts exhibit marked differences in
their spectral extent and spectral peaks.  Some bursts, such as 11A, show
emission across the entire observed band; others, such as 11C, 11H and 11K, show 
emission only at the highest frequencies. Bursts 11G, 11N, and 12A show a
concentration of emission in the lower half of the band. Bursts 11A to 11F
exhibit a spectral peak around 7~GHz, which appears to then shift to lower frequencies in later bursts.   


In order to assess the origin of large- and finer-scale frequency structures 
seen in Figure \ref{fig:collage} and \ref{fig:dynamic_spectra}, we calculated 
the Galactic scintillation bandwidth across the observed band. 
We used the NE2001 model \citep{Cor02}, and estimated a Galactic scattering 
timescale $\tau_{s} = 20\,\mu{\rm s}~{\nu}^{-\alpha}$ towards the direction of 
FRB\,121102 ($\nu$ is the observing frequency in GHz while $\alpha$ is the scaling parameter). 
Assuming a frequency scaling for a Kolmogorov spectrum with scaling parameter between 4 to 4.4 \citep{Cor02,bcc04}, 
we estimate the expected diffractive interstellar scintillation bandwidth as \citep{lm99}
\begin{equation}
    {\Delta}f_{\rm DISS}~=~ \frac{1.16}{2{\pi}{\tau_{\rm s}}}~=~\frac{1.16{\nu}^{[4,4.4]}}{40\pi} ~ {\rm MHz.} 
    \label{eq:diss}
\end{equation}
For the observed band (4.5 to 8\,GHz), the predicted ${\Delta}f_{\rm DISS}$ varies from 7\,MHz to 87\,MHz 
(assuming a scaling parameter of 4.4). Thus, the large-scale frequency structures ($\sim$ GHz wide) are 
likely to be intrinsic to the source and/or due to propagation effects in the source's local environment.

\begin{figure}[h]
\centering
    \begin{center}
    \includegraphics[scale=0.45]{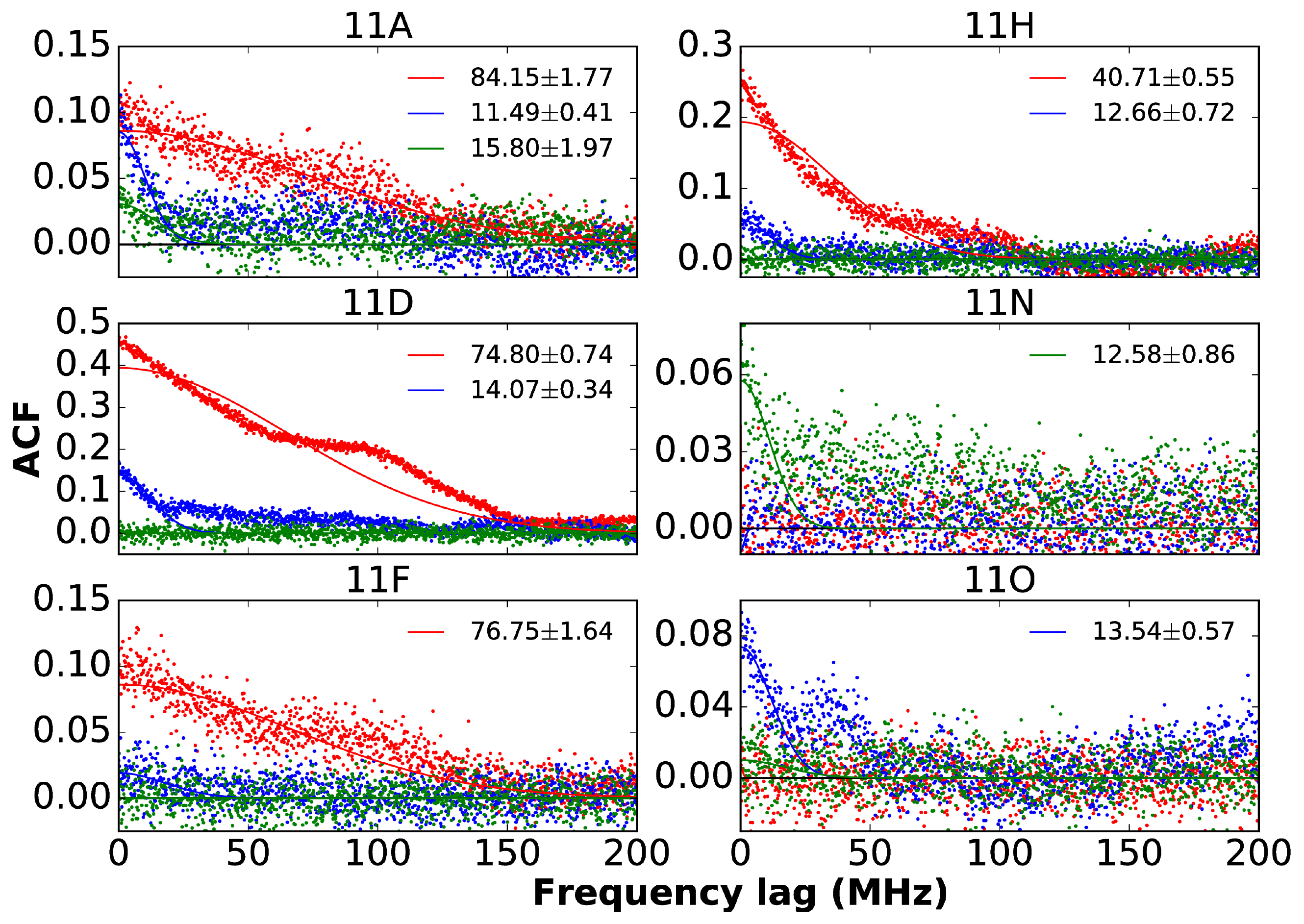}
    \caption{Example ACFs from different subbands of the entire $4.5 - 8$\,GHz 
    bandwidth 
    for six of the FRB\,121102 bursts presented here. 
    The red, blue and green color 
    dots with fitted Gaussian functions are from sub-spectra centered at frequencies 
    of 7500, 6500 and 5200\,MHz, respectively.  The characteristic bandwidths (in 
    MHz) for each subband are shown as inset text with corresponding errors 
    ($1 \sigma$).}
    \label{fig:ACF_plots}
    \end{center}
\end{figure}

To compare these predicted values for ${\Delta}f_{\rm DISS}$ with fine-scale frequency 
structures from the detected bursts, we obtained spectra for each burst by selecting 
an appropriate on-pulse window. Following the procedure of \cite{cwb85} and \cite{cor86}, 
we then computed auto-correlation functions (ACFs) for these spectra (Figure \ref{fig:ACF_plots}). 
As the spectral extent of a few bursts were significantly broad---spanning over 3\,GHz---the ACFs
over the entire spectra produce a multi-featured ACF with multiple widths. To measure
the characteristic bandwidth for a given burst, we divided each spectrum into two or
three parts, depending upon the spectral extent.  We then fitted a Gaussian
function to the ACF to obtain the half width at half maximum (HWHM) as a
characteristic bandwidth for the given sub-spectra for each burst.

\begin{table}[h]
\begin{center}
\small
\begin{tabular}{l c c}
\hline
Burst & Frequency (GHz) &$\Delta{f}$ (MHz) \\
\hline
\hline
\multirow{3}{*}{11A} & 7.595 &  84.2$\pm$5.3 \\
                     & 6.407 &  11.5$\pm$1.2 \\
                     & 5.220 &  15.8$\pm$5.9 \\
\hline               
\multirow{2}{*}{11D} & 7.595 &  74.8$\pm$2.2 \\
                     & 6.407 &  14.1$\pm$1.0 \\
\hline                     
\multirow{3}{*}{11E} & 7.595 &  73.1$\pm$11.7 \\
					 & 6.407 &  19.1$\pm$13.6 \\
					 & 5.220 &  13.9$\pm$11.7 \\
\hline                     
\multirow{2}{*}{11F} & 7.595 &  76.7$\pm$4.9\\
					 & 6.407 &  20.5$\pm$17.4\\  
\hline                     
11G					 & 5.220  &  2.3$\pm$0.7 \\                     
\hline
\multirow{2}{*}{11H} & 7.595 &  40.7$\pm$0.5 \\
					 & 6.407 &  12.7$\pm$0.7 \\
\hline                     
11I 				 & 5.517 &  15.7$\pm$1.9 \\
\hline
\multirow{2}{*}{11J} & 7.298 &  59.6$\pm$7.8 \\
					 & 5.517 &  18.1$\pm$6.7\\
\hline                     
11K 				 & 6.407 &  24.4$\pm$8.8\\
\hline
11N 				 & 5.517 &  15.1$\pm$1.2 \\
\hline
\multirow{2}{*}{11O} & 7.298 &  82.4$\pm$15.1 \\
					 & 5.517 &  17.1$\pm$1.1\\
\hline                     
11Q 				 & 6.407 &  17.0$\pm$0.4 \\
\hline
12A 				 & 5.517 &  13.3$\pm$0.5\\
\hline
\multirow{2}{*}{12B} & 7.595 &  83.1$\pm$4.6 \\
					 & 6.407 &  16.7$\pm$1.2\\
\hline                     
12C 				 & 5.517 &  16.3$\pm$0.9 \\
\hline
\end{tabular}
\caption{Table of characteristic bandwidths for 15 bursts from FRB\,121102 across 4.5 to 8\,GHz. 
The second column lists the center frequency of the sub-spectra used to calculate the characteristic 
bandwidths from ACFs listed in the third column with corresponding errors (3$\sigma$).}
\label{tab:acf_values}
\normalsize
\end{center}
\end{table}

\begin{figure}[h]
    \begin{center}
    \includegraphics[scale=0.8]{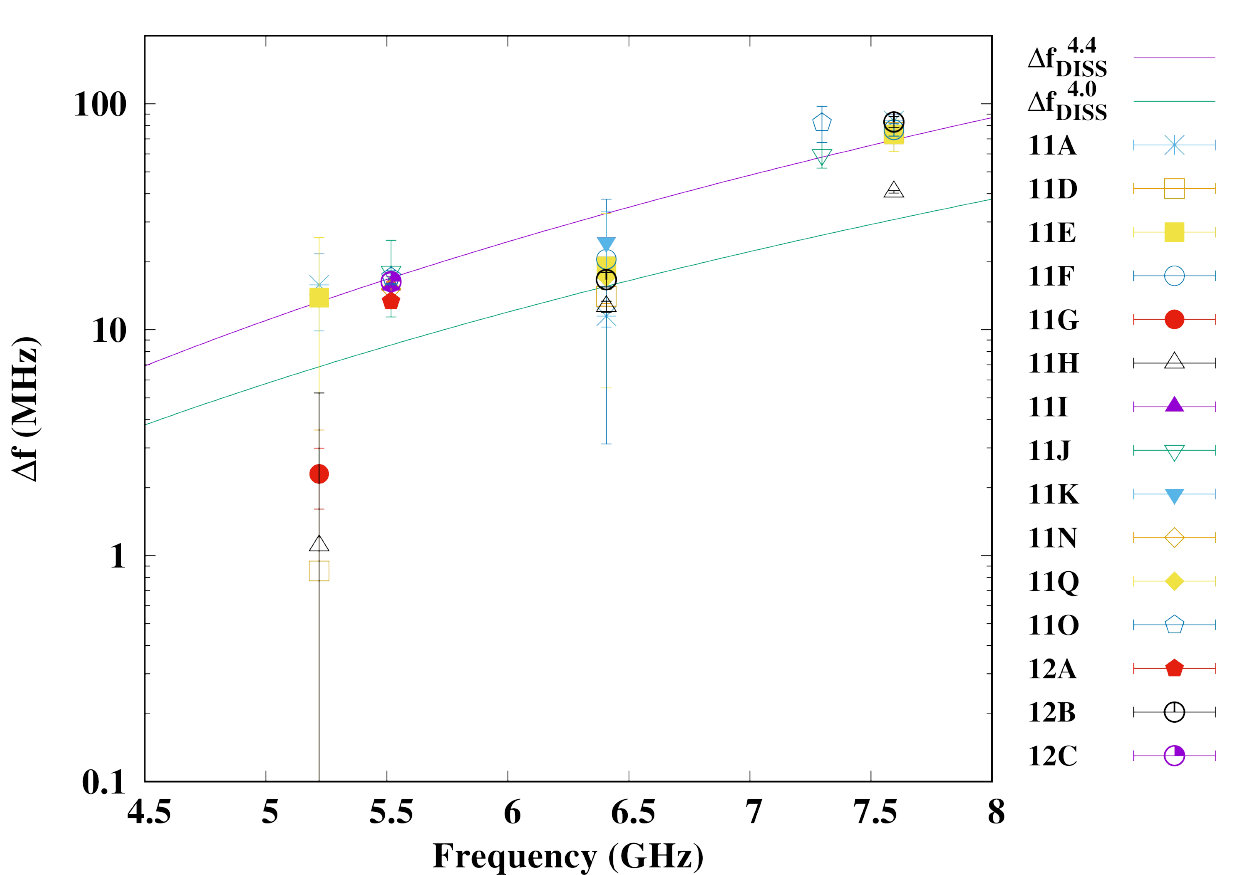}
    \caption{Measured characteristic bandwidth (in MHz) of 15 individual strong bursts, for various subbands, 
    along with the expected ${\Delta}f_{\rm DISS}$ (with two different scaling parameters) 
    from Galactic diffractive scintillations (see Equation \ref{eq:diss}).}
    \label{fig:DISS_bandwidth}
    \end{center} 
\end{figure}

We found that the ACFs showed different characteristic bandwidths in different sub-spectra 
in the 4.5 to 8\,GHz band. Table \ref{tab:acf_values} shows these characteristic bandwidths 
from 15 stronger bursts. The measured characteristic bandwidth follows the 
expected {\itshape ${\Delta}f_{\rm DISS}$} across the 4.5 to 8~GHz band (see Figure \ref{fig:DISS_bandwidth}) 
suggesting that the fine-scale frequency structures seen in each of these bursts
are likely due to Galactic interstellar scintillation, with a few exceptions.
For example, burst 11H does not appear to show characteristic bandwidth matching
with the $\Delta{f}_{\rm DISS}$ at higher frequencies. However, this could be a 
statistical fluctuation given the small number of scintles in the burst.

\section{Discussion}
\label{sect:discussion}


\subsection{Instantaneous Burst Rate}
Our detection of 21 bursts within 60 minutes represents the highest number of bursts 
detected within a short interval (i.e. hour timescale), with 18 bursts
occurring in the first 30 minutes. We did not detect any bursts during the following 
4 hours of observations, which supports the idea that FRB\,121102's bursting behavior 
is episodic. This could be due to intrinsic changes in the emission conditions or due 
to more favorable `plasma lensing' conditions \citep{Cor17} during the first hour, 
potentially enhancing observed burst energies by an order of magnitude. 

Our observations highlight the advantages afforded by wider instantaneous
bandwidth. If our 4~GHz instantaneous bandwidth were halved, we would only have
detected 10 to 12 bursts, mainly as the band-limited spectral structures 
they exhibit would have fallen out of band (Figure~\ref{fig:collage}). 
\cite{Law17} have also pointed out similar limitations of narrow bandwidth observations. 

The spectral peak variation of each burst and apparent band-limited characteristics indicate
that further bursts may be detected by searching for pulses over subbands, particularly 
in observations over large fractional bandwidths. The existence of spectral structure 
may also impact the performance of single-pulse search pipelines, which are generally 
designed to search for broadband pulses.

\subsection{Average spectral and temporal properties across 1 -- 8 GHz}
\begin{figure*}[h]
    \begin{center}
    \subfigure[]{
    \includegraphics[scale=0.8]{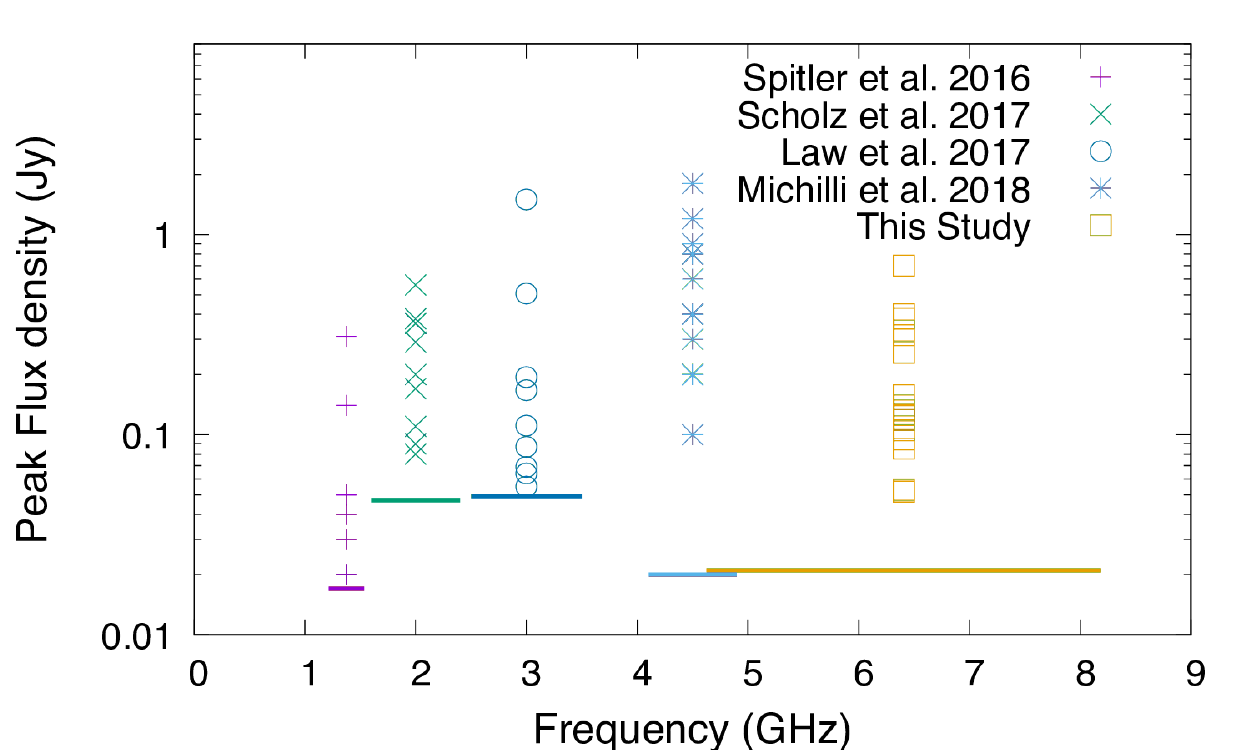}
    \label{fig:freq_vs_flux}
    }
    \subfigure[]{
    \includegraphics[scale=0.8]{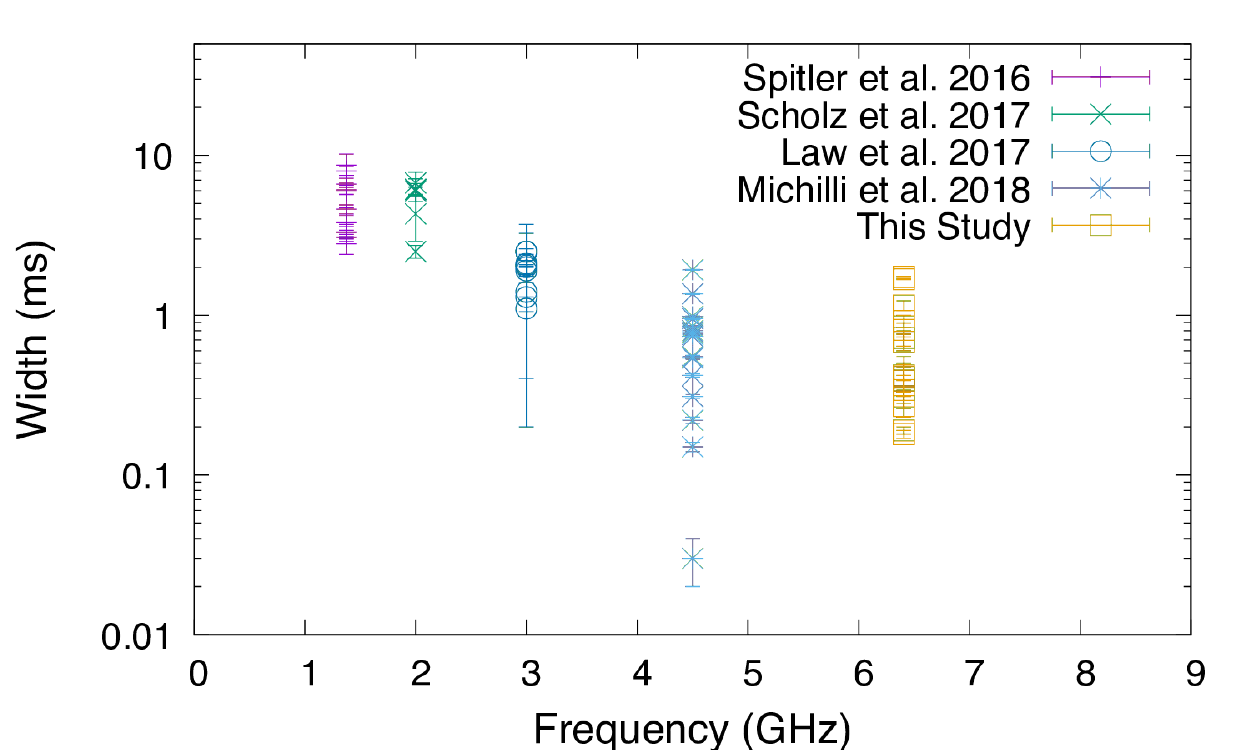}
    \label{fig:freq_vs_width}
    }
    \caption{(a) Average peak flux densities as a function of observed frequency from this and previous studies. 
    The horizontal lines represents the minimum detectable flux density ($\geq$ 6$\sigma$) 
    for a 1-ms wide pulse, with horizontal extent showing the observational frequency bandwidths
    at different frequencies. (b) Observed burst widths as a function of observed frequency from previous studies, 
    indicating that higher frequency bursts are relatively narrower.}
    \end{center}
\end{figure*}

We compared the average peak flux densities of all our bursts with
previously reported observations at various lower frequencies, as shown in Figure \ref{fig:freq_vs_flux}. 
We found that, statistically, the distribution of peak flux densities across frequency is consistent with being flat across $1 - 8$\,GHz. It should be noted that these measurements were obtained from different telescopes at very different epochs which might affect their intrinsic absolute scaling. The different sensitivities of different telescopes at 
various frequencies are also highlighted in Figure \ref{fig:freq_vs_flux}. We also compared the burst widths 
at various frequencies (Figure \ref{fig:freq_vs_width}) and highlight that bursts are relatively narrower 
at higher frequencies ($> 4$\,GHz).

The apparently flat spectrum of FRB\,121102 stands in contrast to the steep spectral indices 
observed for most neutron star emission.  Giant pulses from the Crab pulsar, for instance, 
typically exhibit a very steep spectrum with a power-law index $\alpha = -2.6$ and with a 
steep power-law distribution of rates at a fixed fluence \citep{2017ApJ...851...20M}.  
Ordinary pulsars also show very steep spectra with a mean index $\alpha = -1.4$ \citep{blv13}.  
An important exception is the radio to millimeter spectrum of magnetars.  The Galactic Center magnetar, 
SGR\,J1745$-$2900, has been observed to have a flat spectrum up to 291\,GHz \citep{2017MNRAS.465..242T}.  
The similarity in spectral index between FRB\,121102 and radio magnetars suggests a common 
emission mechanism. It is likely that further high frequency observations with better 
sensitivity, enabling the detection of weaker bursts, may provide better constraints on the steepness of the spectral index. 

If the apparently flat spectral behavior of FRB\,121102 is a common property for other FRBs, 
we suggest that future FRB surveys could be conducted effectively
at higher frequencies, utilizing either fly's eye mode or multiple beams to compensate for a smaller field of view. 
Higher-frequency observations may also have lower terrestrial radio interference and larger instantaneous receiver 
bandwidth --- potentially beneficial for detecting more of these spectrally-limited bursts. Similar suggestions were also made by \cite{Law17}. 

\subsection{Frequency structures}
The large-scale frequency structures are unlikely to be instrumental in 
nature and are likely intrinsic to the progenitor or propagation effects 
imparted in the source's local environment. We note the similarity of these structures to the banded structures seen for the Crab giant pulses (GPs) at similar higher radio frequencies reported by \cite{he07} and \cite{jon10}.  
We also found that burst-to-burst spectral properties change on the order of
tens of seconds.  If bursts from FRB\,121102 have a physical origin similar to 
Crab GPs, then such changes are comparable to similar spectral feature variations 
between Crab GPs, although these manifest at a shorter timescale (few microseconds).

\subsection{High Faraday Rotation Measure}
The high RM found here for FRB\,121102 is almost 500 times more than RMs reported
for any other FRB (e.g. \citealt{Mas15}) and somewhat larger than the RM of the 
Galactic center magnetar SGR\,J1745$-$2900 \citep{Eat13}. Our measured RM is 
about 10\% lower than the RM from bursts detected at $4 - 5$\,GHz, from data obtained 
seven months prior to the observations reported here. This change in the RM is already highlighted in \cite{Michilli:2018}. This is a significant change in the RM and further justifies regular monitoring to 
clarify how the RM varies with time. SGR\,J1745$-$2900 also shows changes in the RM of similar scale over four years of regular monitoring \citep{dep18}. 

The high RM suggests an intense magnetic field, of order 1\,mG, in the
progenitor's environment. As noted in \cite{Michilli:2018}, plausible scenarios
that could produce this RM include: the source being in vicinity of a
intermediate mass or supermassive black-hole, like SGR J1745$-$2900 \citep{Eat13}; 
inside a powerful pulsar wind nebula, or a supernova remnant. 
\cite{piro16} suggested that an expanding supernova shell could also cause the
RM to decrease with time as reported here. However, such changes can also cause 
the corresponding DM to decrease with time by a similar factor, which was not observed. 
Other FRBs might have similar high RMs; however, measuring large 
RMs requires higher frequency resolution at lower frequencies 
(e.g. $\leq 3$\,GHz) to avoid intra-channel depolarization. 
For example, to search for RM up to 10$^{5}$ rad m$^{-2}$, the required channel resolution is of order 
tens of kHz at 1\,GHz. This again highlights the utility of high-frequency observations of 
FRBs as intra-channel depolarization is inversely proportional to observational frequency to the third power. 

\section{Conclusion}
\label{sect:conclusion}
We have reported, for the first time, that FRB\,121102 is active above 5.2\,GHz.  
The 21 bursts detected over 60 minutes represent the highest instantaneous 
burst-rate yet observed.  We have confirmed that bursts from FRB\,121102 are
highly linearly polarized, that the source shows a large RM of
$93589 \pm 118$~rad~m$^{-2}$, and that the diverse and variable spectral and
temporal properties seen at lower frequencies are also exhibited above 5\,GHz. 
As at lower frequencies \citep{Spi16,Sch16,Law17}, we also found that the source 
exhibits large-scale frequency structures that could be intrinsic, imparted 
by its local environment. These structures vary between bursts and can bias 
the estimation of burst properties, such as dispersion measure and pulse width 
as they superimpose to provide enhanced S/N at higher trial DMs. 
We found fine-scale structure consistent with 
Galactic interstellar scintillation.  Future observations of this source will 
help to answer some of the questions yet outstanding, including whether the 
source exhibits any periodicity, at what frequency the apparently flat spectral 
index transitions and how the event rate varies as a function of frequency. 

\section*{Acknowledgements}
VG thanks A. Noutsos for useful comments on the draft. We would like to thank our referee, J. I. Katz 
for a critical review of the paper and for suggesting several improvements to the manuscript. VG would like to acknowledge NSF grant 1407804 and the Marilyn and Watson Alberts SETI Chair funds. CJL is supported by the NSF award 1611606. DM and JWTH acknowledge support from the European Research Council under the European Union's Seventh Framework Programme (FP/2007-2013) / ERC Grant Agreement nr. 337062. JWTH also acknowledges funding from an NWO Vidi fellowship. L.G.S. acknowledges financial support from the ERC Starting Grant BEACON under contract number 279702, as well as the Max Planck Society. MAM and SBS are supported by NSF awards NSF-OIA award number 1458952. SBS, SC, JMC, TJWL, MAM, SMR, and AS are members of the NANOGrav Physics Frontiers Center supported by NSF award PHY-1430284. NG is supported by HOU for her current sabbatical leave. Breakthrough Listen is managed by the Breakthrough Initiatives, sponsored by the Breakthrough Prize Foundation (\href{http://breakthroughinitiatives.org}{breakthroughinitiatives.org}). The Green Bank Observatory is a facility of the National Science Foundation operated under cooperative agreement by Associated Universities, Inc. 

\bibliographystyle{aasjournal}


\end{document}